\documentclass[11pt,a4paper]{article}

\usepackage{amssymb}
\usepackage{amscd}
\usepackage{amsmath}
\begin{document}


\newcounter{mo}
\newcommand{\mo}[1]
{\stepcounter{mo}$^{\bf MO\themo}$%
\footnotetext{\hspace{-3.7mm}$^{\blacksquare\!\blacksquare}$
{\bf MO\themo:~}#1}}

\newcounter{bk}
\newcommand{\bk}[1]
{\stepcounter{bk}$^{\bf BK\thebk}$%
\footnotetext{\hspace{-3.7mm}$^{\blacksquare\!\blacksquare}$
{\bf BK\thebk:~}#1}}


\newcommand{\Si}{\Sigma}
\newcommand{\tr}{{\rm tr}}
\newcommand{\ad}{{\rm ad}}
\newcommand{\Ad}{{\rm Ad}}
\newcommand{\ti}[1]{\tilde{#1}}
\newcommand{\om}{\omega}
\newcommand{\Om}{\Omega}
\newcommand{\de}{\delta}
\newcommand{\al}{\alpha}
\newcommand{\te}{\theta}
\newcommand{\vth}{\vartheta}
\newcommand{\be}{\beta}
\newcommand{\la}{\lambda}
\newcommand{\La}{\Lambda}
\newcommand{\D}{\Delta}
\newcommand{\ve}{\varepsilon}
\newcommand{\ep}{\epsilon}
\newcommand{\vf}{\varphi}
\newcommand{\vfh}{\varphi^\hbar}
\newcommand{\vfe}{\varphi^\eta}
\newcommand{\fh}{\phi^\hbar}
\newcommand{\fe}{\phi^\eta}
\newcommand{\G}{\Gamma}
\newcommand{\ka}{\kappa}
\newcommand{\ip}{\hat{\upsilon}}
\newcommand{\Ip}{\hat{\Upsilon}}
\newcommand{\ga}{\gamma}
\newcommand{\ze}{\zeta}
\newcommand{\si}{\sigma}

\def\hS{{\hat{S}}}

\newcommand{\li}{\lim_{n\rightarrow \infty}}
\def\mapright#1{\smash{
\mathop{\longrightarrow}\limits^{#1}}}

\newcommand{\mat}[4]{\left(\begin{array}{cc}{#1}&{#2}\\{#3}&{#4}
\end{array}\right)}
\newcommand{\thmat}[9]{\left(
\begin{array}{ccc}{#1}&{#2}&{#3}\\{#4}&{#5}&{#6}\\
{#7}&{#8}&{#9}
\end{array}\right)}
\newcommand{\beq}[1]{\begin{equation}\label{#1}}
\newcommand{\eq}{\end{equation}}
\newcommand{\beqn}[1]{\begin{small} \begin{eqnarray}\label{#1}}
\newcommand{\eqn}{\end{eqnarray} \end{small}}
\newcommand{\p}{\partial}
\def\sq2{\sqrt{2}}
\newcommand{\di}{{\rm diag}}
\newcommand{\oh}{\frac{1}{2}}
\newcommand{\su}{{\bf su_2}}
\newcommand{\uo}{{\bf u_1}}
\newcommand{\SL}{{\rm SL}(2,{\mathbb C})}
\newcommand{\GLN}{{\rm GL}(N,{\mathbb C})}
\def\sln{{\rm sl}(N, {\mathbb C})}
\def\sl2{{\rm sl}(2, {\mathbb C})}
\def\SLN{{\rm SL}(N, {\mathbb C})}
\def\SLT{{\rm SL}(2, {\mathbb C})}
\def\PSLN{{\rm PSL}(N, {\mathbb C})}
\newcommand{\PGLN}{{\rm PGL}(N,{\mathbb C})}
\newcommand{\gln}{{\rm gl}(N, {\mathbb C})}
\newcommand{\PSL}{{\rm PSL}_2( {\mathbb Z})}
\def\f1#1{\frac{1}{#1}}
\def\lb{\lfloor}
\def\rb{\rfloor}
\def\sn{{\rm sn}}
\def\cn{{\rm cn}}
\def\dn{{\rm dn}}
\newcommand{\rar}{\rightarrow}
\newcommand{\upar}{\uparrow}
\newcommand{\sm}{\setminus}
\newcommand{\ms}{\mapsto}
\newcommand{\bp}{\bar{\partial}}
\newcommand{\bz}{\bar{z}}
\newcommand{\bw}{\bar{w}}
\newcommand{\bA}{\bar{A}}
\newcommand{\bG}{\bar{G}}
\newcommand{\bL}{\bar{L}}
\newcommand{\btau}{\bar{\tau}}

\newcommand{\tie}{\tilde{e}}
\newcommand{\tial}{\tilde{\alpha}}

\newcommand{\Sh}{\hat{S}}
\newcommand{\vtb}{\theta_{2}}
\newcommand{\vtc}{\theta_{3}}
\newcommand{\vtd}{\theta_{4}}

\def\mC{{\mathbb C}}
\def\mZ{{\mathbb Z}}
\def\mR{{\mathbb R}}
\def\mN{{\mathbb N}}

\def\frak{\mathfrak}
\def\gb{{\frak b}}
\def\gg{{\frak g}}
\def\gp{{\frak p}}
\def\gn{{\frak n}}
\def\gJ{{\frak J}}
\def\gS{{\frak S}}
\def\gL{{\frak L}}
\def\gM{{\frak M}}
\def\gG{{\frak G}}
\def\gE{{\frak E}}
\def\gF{{\frak F}}
\def\gk{{\frak k}}
\def\gK{{\frak K}}
\def\gl{{\frak l}}
\def\gh{{\frak h}}
\def\gu{{\frak u}}
\def\gH{{\frak H}}
\def\gs{{\frak s}}
\def\gt{{\frak t}}
\def\gT{{\frak T}}
\def\gR{{\frak R}}

\def\baal{\bar{\al}}
\def\babe{\bar{\be}}

\def\bfa{{\bf a}}
\def\bfb{{\bf b}}
\def\bfc{{\bf c}}
\def\bfd{{\bf d}}
\def\bfe{{\bf e}}
\def\bff{{\bf f}}
\def\bfg{{\bf g}}
\def\bfm{{\bf m}}
\def\bfn{{\bf n}}
\def\bfp{{\bf p}}
\def\bfu{{\bf u}}
\def\bfv{{\bf v}}
\def\bfr{{\bf r}}
\def\bfs{{\bf s}}
\def\bft{{\bf t}}
\def\bfx{{\bf x}}
\def\bfy{{\bf y}}
\def\bfM{{\bf M}}
\def\bfR{{\bf R}}
\def\bfC{{\bf C}}
\def\bfP{{\bf P}}
\def\bfq{{\bf q}}
\def\bfS{{\bf S}}
\def\bfJ{{\bf J}}
\def\bfI{{\bf I}}
\def\bfX{{\bf X}}
\def\bfT{{\bf T}}
\def\bfz{{\bf z}}
\def\bfnu{{\bf \nu}}
\def\bfA{{\bf A}}
\def\bfB{{\bf B}}
\def\bfE{{\bf E}}
\def\bfF{{\bf F}}
\def\bfsi{{\bf \sigma}}
\def\bfU{{\bf U}}
\def\bfso{{\bf so}}

\def\clA{\mathcal{A}}
\def\clC{\mathcal{C}}
\def\clD{\mathcal{D}}
\def\clE{\mathcal{E}}
\def\clG{\mathcal{G}}
\def\clF{\mathcal{F}}
\def\clR{\mathcal{R}}
\def\clU{\mathcal{U}}
\def\clT{\mathcal{T}}
\def\clO{\mathcal{O}}
\def\clH{\mathcal{H}}
\def\clK{\mathcal{K}}
\def\clJ{\mathcal{J}}
\def\clI{\mathcal{I}}
\def\clL{\mathcal{L}}
\def\clM{\mathcal{M}}
\def\clN{\mathcal{N}}
\def\clP{\mathcal{P}}
\def\clQ{\mathcal{Q}}
\def\clS{\mathcal{S}}
\def\clV{\mathcal{V}}
\def\clW{\mathcal{W}}
\def\clZ{\mathcal{Z}}

\def\baf{{\bf f_4}}
\def\bae{{\bf e_6}}
\def\ble{{\bf e_7}}
\def\bag2{{\bf g_2}}
\def\bas8{{\bf so(8)}}
\def\baso{{\bf so(n)}}

\def\sr2{\sqrt{2}}
\newcommand{\ran}{\rangle}
\newcommand{\lan}{\langle}
\def\f1#1{\frac{1}{#1}}
\def\lb{\lfloor}
\def\rb{\rfloor}
\newcommand{\slim}[2]{\sum\limits_{#1}^{#2}}

\renewcommand{\theequation}{\thesection.\arabic{equation}}
\newtheorem{predl}{Proposition}[section]
\newtheorem{defi}{Definition}[section]
\newtheorem{rem}{Remark}[section]
\newtheorem{cor}{Corollary}[section]
\newtheorem{lem}{Lemma}[section]
\newtheorem{theor}{Theorem}[section]


\def\theequation{\arabic{equation}}



\vspace{0.1in}
\begin{flushright}
 ITEP/TH-18/17\\
 IITP/TH-11/17
\end{flushright}
\vspace{2mm}

\begin{center}
{\large{\bf Calogero-Sutherland system with two types interacting
spins
 }
}\\
\vspace{10mm} {S. Kharchev$^{\flat\,\S}$ {{ A.
Levin}$^{\,\natural\,\,\flat}$ { M. Olshanetsky}$^{\,\flat\,\S}$
 { A. Zotov}$^{\,\diamondsuit\,\flat}$}\\ \vspace{7mm}
 \vspace{2mm} $^\flat$ - {\sf Institute of Theoretical and Experimental Physics,
  Moscow, 117218, Russia}\\
 \vspace{2mm}$^\natural$ - {\sf International Laboratory for Mirror Symmetry and Automorphic Forms,\\
  Mathematics Department of NRU HSE,
 Usacheva str. 6,  Moscow, 119048, Russia}\\
 \vspace{2mm} $^\diamondsuit$ -
 {\sf Steklov Mathematical Institute RAS, Gubkina str. 8, Moscow, 119991,  Russia
 }\\
 \vspace{2mm}$^\S$ - {\sf Institute for Information Transmission Problems RAS (Kharkevich Institute),
 \\  Bolshoy Karetny per. 19, Moscow, 127994,  Russia}\\
}
 \vspace{4mm}
 {\footnotesize Emails: kharchev@itep.ru, alevin2@hse.ru, olshanet@itep.ru,
 zotov@mi.ras.ru}\\
 \vspace{4mm}
\textbf{Abstract}
\end{center}

We consider the classical Calogero-Sutherland system with two types
of interacting spin variables. It can be reduced to the standard
Calogero-Sutherland system, when one of the spin variables vanishes.
 We describe the model in the Hitchin approach and
 prove complete integrability of the system by constructing the Lax
 pair and the classical $r$-matrix with the spectral parameter on a singular curve.

\vspace{10mm}


\noindent \paragraph{Introduction.}

The Calogero-Sutherland (CS) model \cite{Ca,Su,OP} describes
one-dimensional system of  interacting pairwise particles through
long range potentials. It has a lot of applications, in particular
with the quantum Hall effect \cite{AI}, matrix models \cite{A}, and
orthogonal polynomials \cite{BF}.
%
In this paper we consider the classical case. The classical CS model
is an integrable system in the Liouville sense. Moreover, it remains
integrable if one adds the so-called spin variables. The resulting
system has form of the Euler-Arnold SL$(N)$ top with the inertia
tensor depending on the positions of interacting particles
\cite{GH}.

Denote the coordinates of the particles $\bfu=(u_1,\ldots,u_N)$,
their momenta $\bfv=(v_1,\ldots,v_N)$ and the spin variables
$\{S_{jk}\}$ arranged into matrix ${S}=\sum_{ij}^N E_{ij} S_{ij}$
(here $\{E_{ij}\}$ is the standard basis in Mat$(N)$, i.e.
$\left(E_{ij}\right)_{ab}=\delta_{ia}\delta_{bj}$ ). The latter is
an element of the Lie algebra ${\rm sl}(N)$. The spin CS model is
described by the Hamiltonian
  \beq{i0}
H^{CS}=\oh\sum_{j=1}^Nv_j^2-\sum_{j<
k}\frac{S_{jk}S_{kj}}{\sinh^{2}(u_j-u_k)}\,.
  \eq
The Poisson brackets between positions of particles and momenta are
canonical $\{v_k,u_j\}=\de_{jk}$, while the Poisson structure for
$\{S_{jk},S_{mn}\}$ is given by the Dirac brackets. They can be
obtained starting from the Lie-Poisson brackets on the Lie coalgebra
${\rm sl}^*(N)$ after imposing constraints ${S}_{diag}=0$ (and some
gauge fixation) resulting from the
coadjoint action of the diagonal subgroup of SL$(N)$ on the spin
variables ${S}$. The case when $S\in{\rm so}(N)$ is know as well
\cite{BAB}. Some further generalizations can be found in
\cite{Feher}.

\noindent \paragraph{Summary.} Our generalization of (\ref{i0}) is
as follows:
  \beq{i1}
H=\oh\sum_{j=1}^Nv_j^2+\sum_{j< k}\frac{ S_{jk}^2 +T_{jk}^2-2
S_{jk}T_{jk}\cosh(u_j-u_k)} {\sinh^{2}(u_j-u_k)}\,,
  \eq where
$S_{ij}, T_{ij}$ are elements of antisymmetric matrices ${S}$ and
${T}$ ($S_{jk}=-S_{kj}$, $T_{jk}=-T_{kj}$) with the Lie-Poisson
brackets on the direct sum of two Lie coalgebras ${\rm
so}^*(N)\oplus{\rm so}^*(N)$:
  \beq{io3}
\begin{array}{c}
 \displaystyle{
 \{S_{ij},S_{kl}\}=
 -\frac12\,\left( S_{il}\delta_{kj}-S_{kj}\delta_{il}-S_{ik}\delta_{lj}+S_{lj}\delta_{ik}
 \right)\,,
 }
  \\ \ \\
 \displaystyle{
 \{T_{ij},T_{kl}\}=
 \frac12\,\left( T_{il}\delta_{kj}-T_{kj}\delta_{il}-T_{ik}\delta_{lj}+T_{lj}\delta_{ik}
 \right)\,,}\\ \ \\
 \displaystyle{\quad\quad \{S_{ij},T_{kl}\}=0\,.
 }
 \end{array}
 \eq
 The phase space ${\mathbb R}^{2N-2}\times {\mathcal O}_{{\rm SO}(N)}\times {\mathcal O}_{{\rm SO}(N)}$ consists
 of ${\mathbb R}^{2N-2}$ parameterized by momenta and positions of
 $N$ particles in the center of mass frame and two coadjoint
 orbits ${\mathcal O}_{{\rm SO}(N)}$.  Each orbit is obtained from
${\rm so}^*(N)$ by fixation of the Casimir functions (i.e. the
eigenvalues of matrices $S$ and $T$). The Poisson structure
(\ref{io3}) keeps the same form on ${\mathcal O}_{{\rm SO}(N)}\times
{\mathcal O}_{{\rm SO}(N)}$. The dimension of generic SO$(N)$ orbit
equals\footnote{$[x]$ stands for integer part of $x$.}
$(1/2)(N^2-N)-[N/2]$.
  Therefore, the dimension of the total phase space is   $(N-1)(N+2)-2[N/2]$.

  For $N=2$ the Lie
algebra so(2) is commutative and the  spin variables are fixed. In
this case we obtain from (\ref{i1}) the Hamiltonian with two
constants
  \beq{i4}
 H=\frac{v^2}{2}+\frac{
m_1^2 +m_2^2-2 m_1m_2\cosh(2u)} {\sinh^{2}(2u)}\,,
  \eq
which reproduces the CS model of  the BC$_1$ type \cite{OP}.

We prove that there exists $(1/2)(N-1)(N+2)-[N/2]$ independent
integrals of motion in involution, and the Hamiltonian (\ref{i1}) is
one of them. To this end we construct the Lax pair and the classical
$r$-matrix. Similarly to so$(N)$ version of (\ref{i0}) \cite{BAB}
(and in contrast to the sl$(N)$ spin CS system) the $M$-operator can
be explicitly constructed because
 the spin variables ${S},{T}$ are skew-symmetric matrices,  and the additional reduction is not needed.

We derive the generalized CS system (GCS) using the Hitchin approach
\cite{Hi,Hi1}. The Lax operator of integrable system satisfies the
Hitchin equations. They come  from the self-duality equations in
four dimension after their reduction to two dimensional Riemann
surface. Namely, instead of $\mR^4$ one consider the
four-dimensional space $\mR^2\times\Si$, where $\Si$ plays the role
of the base spectral curve\footnote{The spectral curve of integrable
system is defined as characteristic equation for the Lax matrix. It
is a branched covering of the base spectral curve, where the
spectral parameter lives.}. The field content of the Hitchin system
comes from the four-dimensional vector-potentials independent on the
first two coordinates. One of them is the Higgs field that plays the role of
the Lax operator of the integrable system.
The coordinates of particles describe the moduli of solutions of
the Hitchin equations, while the spin  variables are the residues of the
Higgs fields at the singular points.
 On Fig.1 and Fig.2 (see the last page) the base spectral curves
$\Si$ of CS and GCS systems are depicted. The Hitchin systems on
singular curve (and, in particular, CS system) were studied
previously in \cite{Ne,TC}.

Another important ingredient of the our construction is the
so-called quasi-compact structure of the gauge group. It means that
the gauge transformations at the singular points on the base spectral curve
are reduced to the unitary group\footnote{In the standard approach
to the Hitchin systems the gauge group may have the quasi-parabolic
structure, i.e. the gauge group is reduced at singular points to the
triangular subgroup.}. We will come to this structure in relation to
integrable systems elsewhere. As a result the spin variables become elements
of the unitary algebra su(N).
To come to the integrable case we further reduce them to the orthogonal algebra so(N).


\noindent \paragraph{Hitchin system structure.}

The described system is the Hitchin system over a singular curve.
\\
\noindent
\emph{The base spectral curve:}\\
The curve is defined in the following way. Consider two rational
curves $\Si_\al\sim\mC P^1$, $(\al=1,2)$ with corresponding
holomorphic coordinates $z_1\,,z_2\in\mC_{1,2}$ Fig.2. They are
glued at the points $z_1^0=z_2^0=0$ and
$z_1^\infty=z_2^\infty=\infty$.  The singular curve is
$\Si=\Si_1\cup\Si_2$ with this identification.

\noindent
\emph{The field content:}\\
1. The anti-holomorphic vector potentials $\bA_\al(z_\al,\bz_\al)$
on the components $\Si_\al$
taking values in the Lie algebra sl$(N,\mC)$;\\
2. The Higgs fields $\Phi_\al(z_\al,\bz_\al)$ are holomorphic
1-forms on $\Si_\al\setminus\{0,1,\infty\}$ taking values in the Lie
algebra sl$(N,\mC)$. They have simple poles at $z_\al=0,1,\infty$
with the definite residues. Let $T$ be an element of the Lie algebra
su$(N)$, $\eta\in$sl$(N,\mC)$, $h\in$SU$(N)$ and $g\in\SLN$. Then we
assume that
  \beq{mmc}
  \begin{array}{lll}
   1.& Res\,\Phi_\al(z_\al=1)|_{su(N)}=0\,,&
   \\
&z_\al=0 & z_\al=\infty \\
2.&  Res\,\Phi_1|_{su(N)}=T\,, & Res\,\Phi_1|_{su(N)}=\eta|_{su(N)}\,, \\
3.&  Res\,\Phi_2|_{su(N)}=hT h^{-1}\,, & Res\,\Phi_2|_{su(N)}=g\eta
g^{-1}|_{su(N)}\,.
 \end{array}
 \eq
These variables $\bA_\al,\Phi_\al$ form \emph{the Higgs bundle}
$\clH_\SLN(\Si)$
over the curve $\Si$.\\
\noindent
\emph{Symplectic structure and symplectic reduction:}\\
 $\clH_\SLN(\Si)$ can be considered as an infinite-dimensional symplectic manifold.
 It is equipped with the symplectic form
 \beq{2.2} \om=\sum_{\al=1,2}\int_{\Si_\al}
D\lan\Phi_\al,D\bA_\al\ran+
D\int_{\Si_1}(\de(z_1,\bz_1)|_{z_1=\infty}\lan\eta,g^{-1}D
g\ran+\de(z_1,\bz_1)|_{z_1=0} \lan T,h^{-1}D h\ran)  \eq
 $$
+D\int_{\Si_2}(\de(z_2,\bz_2)|_{z_2=\infty}\lan\eta,
g^{-1}Dg\ran+\de(z_2,\bz_2)|_{z_2=0} \lan T, h^{-1}D h\ran)\,,
 $$
where $\lan\,,\,\ran$ is the Killing form on sl$(N,\mC)$.
The gauge symmetries of the system are symplectic transformations
preserving (\ref{2.2}). They are formed by the
 pair of smooth maps $f_\al\in\Si_\al\in C^\infty(\Si_\al)\to \SLN$
 such that at the fixed point
  \beq{3.2}
  f_\al(z_\al,\bz_\al)|_{z_\al=0,1,\infty}\in SU(N)\,.
    \eq
The gauge action on the dynamical variables is
  \beq{gt2}
\bp_\al+\bA_\al\to f_\al(\bp_{\al}+\bA_\al) f^{-1}_\al\,,
  \eq
 \beq{gt1}
g\to f_2(\infty)g
f_1^{-1}(\infty)\,,~f_\al(\infty)=f_\al(z_\al,\bz_\al)|_{z_\al=\infty}\,,
 \eq  \beq{gt3} h\to f_2(0) h
f_1^{-1}(0)\,,~f_\al(0)=f_\al(z_\al,\bz_\al)|_{z_\al=0} \,,  \eq
 $$
\Phi_\al\to f_\al\Phi_\al f^{-1}_\al\,,~~T\to f_1 T
f^{-1}_1|_{z_1=0}\,, ~~\eta\to f_1\eta f^{-1}_1|_{z_1=\infty}\,.
 $$
where, according with (\ref{3.2}), in (\ref{gt1})
$f_1,f_2\in$SU$(N)$.
The gauge transformations generate the moment maps
  \beq{m10}
 \mu_1=\p_{\bz_1}\Phi_1+\de(z_1,\bz_1)|_{z_1=\infty}\eta|_{su(N)}+
 \de(z_1,\bz_1)|_{z_1=0}T\,,
 \eq
   \beq{m20}
\mu_2=\p_{\bz_2}\Phi_2+\de(z_2,\bz_2)|_{z_2=\infty}(g\eta
g^{-1})|_{su(N)}+ \de(z_2,\bz_2)|_{z_2=0}S\,,~(S=hTh^{-1})\,.
  \eq
We put the Gauss law constraints $\mu_1=0\,,~~\mu_2=0$. This means
that the Higgs fields $\Phi_\al (z_\al,\bz_\al)$ are meromorphic on
$\Si_\al$ and have simple unitary poles at $z_\al=0,1,\infty$ with
the prescribed residues.
The generic configuration of the vector potentials $\bA_\al$ can be
gauged away ( $\bA_\al=0$) by the gauge group action (\ref{gt2}).
The residual gauge transformations $\clG^{res}_1$ are only constants
$\clG^{res}_1 =\{f_\al\in SU(N)\}$.

For generic $g$ the transformation (\ref{gt1}) allows one to
diagonalize $g$
  \beq{ms1}
   f_2(\infty)g
f_1^{-1}(\infty)=\exp(\bfu)\,,~~\bfu=\di(u_1,\ldots,u_N)\,,~~u_j\in\mR\,.
 \eq
  Thus, the variables $(h,T,\bfu, \bfv)$ describe the reduced phase
space, where $\bfv$ is a variable dual $\bfu$ coming from diagonal
part of $\Phi_\al$. In this way after reduction we come to the
finite-dimensional phase space $\ti\clR^{red}$. It has dimension
$\dim\,\ti\clR^{red}=2(N^2-1)+2(N-1)$.

The gauge transformations are defined up to the action from the
right
 by the final residual gauge transformation $\clG^{res}_2$ preserving the diagonal form of $g$:
  $\clG^{res}_2=
f_\al(\infty)\to f_\al(\infty)s^{-1}\,,$ $s\in
\clG^{res}_2=\clT\rtimes W$, where $W$ is the Weyl group and $\clT$
is the Cartan torus in SU$(N)$.
It implies that:\\
1. $\bfu$ can be ordered as $u_1>u_2>\ldots u_N$.\\
2. The element $h$ in (\ref{ms1}) is defined up to the action $h\to
shs^{-1}$, $s\in \clG^{res}_2=\clT\rtimes W$. Imposing the
corresponding constraints we come to the symplectic quotient
 $\clR^{red}=\ti\clR^{red}//(\clT\rtimes W)$, $\dim\,\clR^{red}=2(N^2-1)$.



\noindent \paragraph{Lax matrix.}

The gauge transformed Higgs fields $\Phi_\al= f_\al L_\al
f^{-1}_\al$, where $f_\al$ is such that
$f_\al(z_\al)|_{z_\al=\infty}$ diagonalize $g$, will play the role
of the Lax operators. The following form of the Lax operators has
the correct structure of poles
  \beq{la} L_1(z_1)=\frac{T}{z_1}+\frac{T+\eta}{1-z_1}\,,~~
L_2(z_2)=\frac{S}{z_2}+\frac{S+g\eta g^{-1}}{1-z_2}\,,~~(S=hT
h^{-1})
  \eq
   if
 $$
\left\{\begin{array}{c}
         T+\eta|_{su(N)}=0\,, \\
         ~S+(g\eta g^{-1})|_{su(N)}=0 \,,
       \end{array}
       \right.
 $$
 cf. (\ref{mmc}). The solution of these equations assumes the form:
 \beq{eta}
 \eta=P+{S}_{diag}+{T}_{diag}+X\,,~~P=\di\,\bfv=\di(v_1,\ldots,v_N)\,,
 \eq
   \beq{13}
 X_{jk}=\frac{S_{jk}-T_{jk} \exp(-u_{jk})}{2\sinh(u_{jk})}\,,~~
  S_{kj}=-\bar S_{jk}\,,~~
 u_{jk}=u_j-u_k\,,~j<k\,.
   \eq

The integrals of motion $I_{lk}$ ($k=0,\dots l$, $l=2,\dots,N$)
 come from the expansion
 \beq{im}
  \lan(T+\eta)^{l}\ran= \lan T^{l}\ran+l\lan
T^{l-1}\eta\ran+\ldots+\lan\eta^{l}\ran\,,
  \eq
  \beq{im1}
I_{lk}=\lan T^{l-k}\eta^k\ran\,, ~~(k=0,\ldots,l)\,.
 \eq
 In
particular, $H$ (\ref{i1}) is $\oh I_{2,2}$. Notice that the
integrals $I_{l,0}$ are the Casimir function of the Lie-Poisson
algebra
 on sl$(N,\mC)^*$.
 Excluding the number of $l$ Casimir functions we obtain the number of integrals
 \beq{ni}
  \clN_G=\sum_{l=2}^N(l+1)-(N-1)=\oh\,(N-1)(N+2)\,.
   \eq
 Evidently, all integrals are functionally independent. But
for the complete integrability in su$(N)$ case we need
$\clN_G=\oh\dim_{\mR}\,(\clR^{red})=N^2-1$, so we have less
integrals of motion than needed.

\paragraph{SO$(N)$ model.} To go around it we pass from the Lie algebra
 su$(N)$ to so$(N)$ for the spin variables ${S},{T}$.
It means that $T_{jk},\,S_{jk}\in\mR$, $S_{jk}=-S_{kj}$,
$T_{jk}=-T_{kj}$ (${S}_{diag}={T}_{diag}=0$) and the eigenvalues of
$S,T$ are fixed. As we explained after (\ref{io3}), the dimension of
the phase space $\dim_{\mR}\,(\clR^{red})=(N-1)(N+2)-2[N/2]$ in this
case, and the number of integrals $\clN_G$  (\ref{ni}) is even more
than needed. It happens because in SO$(N)$ case the Hamiltonians
$I_{l1}$ from (\ref{im1}) for even values of $l$ (in the interval
$2\leq l\leq N$) turn into the Casimir functions (of matrix $T$).
This follows from the skew-symmetry of matrices $S,T$ and can be
verified by direct substitution. Finally, the number of integrals of
motion $\clN_G\,'=\clN_G-[N/2]$ is equal to the half of dimension of
the phase space:
$\oh\dim_{\mR}\,(\clR^{red})=\clN_G\,'=\oh(N-1)(N+2)-[N/2]$.

Let us define the Lax equation on one of the components $\Si_\al$ of
the base spectral curve $\Si$.
Consider the following Lax pair:
 \beq{q032}
 L_{ij}(z)=\delta_{ij}v_i+(1-\delta_{ij})\left(\frac{S_{ij}}{\sinh(u_{ij})}-(\coth(u_{ij})+\coth(z))T_{ij}\right)
 =\ti\eta_{ij}-(1+\coth(z))T_{ij}\,,
 \eq
  \beq{q02}
 M_{ij}=(1-\delta_{ij})\frac{T_{ij}-S_{ij}\cosh(u_{ij})}{\sinh^2(u_{ij})}\,.
  \eq
The Lax matrix (\ref{q032}) is related to $L_1(z_1)$ (\ref{la}) by
$L(z)=(1-z_1)L_1(z_1)$, $\coth(z)=-(1+z_1)/z_1$ with simultaneous
rescaling ${S}\rightarrow 2{S}$ and ${T}\rightarrow 2{T}$ (i.e.
$\ti\eta=P+2X$ in contrast to $\eta=P+X$). The latter is just a
convenient way to avoid unnecessary numerical coefficients.
Finally, the Lax equation
  \beq{q03}
 {\dot L}(z):=\{H,L(z)\}=[L(z),M]
  \eq
with the Lax pair (\ref{q032}), (\ref{q02}) provides equations of
motion generated by the Hamiltonian (\ref{i1}) and the Poisson
structure given by (\ref{io3}) together with the canonical Poisson
brackets $\{v_j,u_k\}=\delta_{jk}$. Indeed, the Lax equation
(\ref{q03}) is equivalent to
 \beq{q04}
 {\dot {\ti\eta}}=[\ti\eta,M]\,,\quad\hbox{where}\quad
  \ti\eta_{ij}=\delta_{ij}v_i+(1-\delta_{ij})\frac{S_{ij}-T_{ij}\exp{(-u_{ij}})}{\sinh(u_{ij})}
 \eq
 and
 \beq{q05}
 {\dot T}=[T,M]\,.
 \eq
From (\ref{q05}) we have
  \beq{q06}
 {\dot T}_{ij}=\sum\limits_{k\neq i,j}\
 T_{ik}T_{kj}\left( \frac{1}{\sinh^2(u_{kj})}-\frac{1}{\sinh^2(u_{ik})}
 \right) -T_{ik}S_{kj}\frac{\cosh(u_{kj})}{\sinh^2(u_{kj})}+S_{ik}T_{kj}\frac{\cosh(u_{ik})}{\sinh^2(u_{ik})}
  \eq
Substitution of (\ref{q06}) into (\ref{q05}) provides equations of
motion for $S$ and $\bfv,\bfu$ variables:
  \beq{q07}
 \displaystyle{
 {\dot u}_i=v_i\,,\quad {\dot v}_i=\sum\limits_{k\neq i}
 2\frac{\cosh(u_{ik})}{\sinh^3(u_{ik})}(S_{ik}^2+T_{ik}^2)-2\frac{\cosh^2(u_{ik})+1}{\sinh^3(u_{ik})}S_{ik}T_{ik}\,,
 }
  \eq
  \beq{q08}
 {\dot S}_{ij}=\sum\limits_{k\neq i,j}\
 S_{ik}S_{kj}\left( \frac{1}{\sinh^2(u_{ik})}-\frac{1}{\sinh^2(u_{kj})}
 \right) +S_{ik}T_{kj}\frac{\cosh(u_{kj})}{\sinh^2(u_{kj})}-
 T_{ik}S_{kj}\frac{\cosh(u_{ik})}{\sinh^2(u_{ik})}\,.
  \eq
It is straightforward to verify that (\ref{q06}),(\ref{q07}) and
(\ref{q08}) are generated by the Hamiltonian (\ref{i1}).


\noindent \paragraph{Classical r-matrix.}
The  $r$-matrix structure is given by
  \beq{q15}
 \displaystyle{
 \{L_1(z),L_2(w)\}=[L_1(z),r_{12}(z,w)]-[L_2(w),r_{21}(w,z)]
 }
  \eq
with $so(N)$ $r$-matrix found in \cite{BAB} for SO$(N)$ spin
Calogero model:
 \beq{q16}
 \begin{array}{c}
 \displaystyle{
 r_{12}(z,w)=\frac12\,(\coth(z-w)+\coth(z+w))\sum\limits_{i}E_{ii}\otimes
 E_{ii}+
 }
 \\ \ \\
 \displaystyle{
 \frac12\,\sum\limits_{i\neq j} E_{ij}\otimes E_{ji} (\coth(z-w)+\coth(u_{ij}))
 +\frac12\,\sum\limits_{i\neq j} E_{ij}\otimes E_{ij} (\coth(z+w)+\coth(u_{ij}))
 }
 \end{array}
 \eq
 The proof of (\ref{q15}) is also straightforward.
 Being written in the form (\ref{q15}) the Poisson brackets  provide the involutivity of the integrals of motion (\ref{im1}).
 One can also verify
that $r$-matrix (\ref{q16}) provides $M$-matrix (\ref{q02}) via
  \beq{q17}
 \displaystyle{
 \tr_2(r_{12}(z,w)L_2(w))=\frac12\,(\coth(z-w)+\coth(z+w))\,L(z)-M\,.
 }
  \eq

\noindent \paragraph{Acknowledgements} This work was funded by the
Russian Science Foundation (RSCF) grant 16-12-10344.


\vspace{5mm}


\unitlength 1mm 
\linethickness{0.4pt}
\ifx\plotpoint\undefined\newsavebox{\plotpoint}\fi 
\begin{picture}(116.029,123.97)(0,0)
\put(115.69,100){\line(0,1){1.046}}
\put(115.667,101.046){\line(0,1){1.044}}
\put(115.599,102.09){\line(0,1){1.04}}
\multiput(115.485,103.13)(-.031845,.206811){5}{\line(0,1){.206811}}
\multiput(115.326,104.164)(-.029171,.146589){7}{\line(0,1){.146589}}
\multiput(115.121,105.19)(-.031097,.127029){8}{\line(0,1){.127029}}
\multiput(114.873,106.206)(-.032543,.111601){9}{\line(0,1){.111601}}
\multiput(114.58,107.211)(-.033644,.099067){10}{\line(0,1){.099067}}
\multiput(114.243,108.202)(-.031612,.081254){12}{\line(0,1){.081254}}
\multiput(113.864,109.177)(-.0324258,.0736588){13}{\line(0,1){.0736588}}
\multiput(113.442,110.134)(-.0330657,.0670184){14}{\line(0,1){.0670184}}
\multiput(112.979,111.072)(-.0335615,.0611442){15}{\line(0,1){.0611442}}
\multiput(112.476,111.99)(-.0319391,.0526071){17}{\line(0,1){.0526071}}
\multiput(111.933,112.884)(-.0323041,.0483208){18}{\line(0,1){.0483208}}
\multiput(111.352,113.754)(-.0325724,.0443986){19}{\line(0,1){.0443986}}
\multiput(110.733,114.597)(-.0327549,.0407881){20}{\line(0,1){.0407881}}
\multiput(110.078,115.413)(-.0328605,.0374476){21}{\line(0,1){.0374476}}
\multiput(109.388,116.199)(-.0328968,.0343426){22}{\line(0,1){.0343426}}
\multiput(108.664,116.955)(-.0343641,.0328743){22}{\line(-1,0){.0343641}}
\multiput(107.908,117.678)(-.0374691,.032836){21}{\line(-1,0){.0374691}}
\multiput(107.121,118.368)(-.0408096,.0327281){20}{\line(-1,0){.0408096}}
\multiput(106.305,119.022)(-.0444199,.0325433){19}{\line(-1,0){.0444199}}
\multiput(105.461,119.641)(-.048342,.0322725){18}{\line(-1,0){.048342}}
\multiput(104.591,120.221)(-.052628,.0319047){17}{\line(-1,0){.052628}}
\multiput(103.696,120.764)(-.0611661,.0335214){15}{\line(-1,0){.0611661}}
\multiput(102.778,121.267)(-.06704,.0330218){14}{\line(-1,0){.06704}}
\multiput(101.84,121.729)(-.07368,.0323776){13}{\line(-1,0){.07368}}
\multiput(100.882,122.15)(-.081275,.031559){12}{\line(-1,0){.081275}}
\multiput(99.907,122.529)(-.099089,.033579){10}{\line(-1,0){.099089}}
\multiput(98.916,122.864)(-.111622,.03247){9}{\line(-1,0){.111622}}
\multiput(97.911,123.157)(-.12705,.031014){8}{\line(-1,0){.12705}}
\multiput(96.895,123.405)(-.146608,.029075){7}{\line(-1,0){.146608}}
\multiput(95.869,123.608)(-.206832,.031709){5}{\line(-1,0){.206832}}
\put(94.834,123.767){\line(-1,0){1.0401}}
\put(93.794,123.88){\line(-1,0){1.044}}
\put(92.75,123.948){\line(-1,0){2.092}}
\put(90.658,123.946){\line(-1,0){1.044}}
\put(89.614,123.877){\line(-1,0){1.0399}}
\multiput(88.574,123.763)(-.20679,-.03198){5}{\line(-1,0){.20679}}
\multiput(87.54,123.603)(-.14657,-.029267){7}{\line(-1,0){.14657}}
\multiput(86.514,123.398)(-.127009,-.03118){8}{\line(-1,0){.127009}}
\multiput(85.498,123.148)(-.11158,-.032616){9}{\line(-1,0){.11158}}
\multiput(84.494,122.855)(-.099045,-.033709){10}{\line(-1,0){.099045}}
\multiput(83.504,122.518)(-.081233,-.031666){12}{\line(-1,0){.081233}}
\multiput(82.529,122.138)(-.0736375,-.032474){13}{\line(-1,0){.0736375}}
\multiput(81.572,121.716)(-.0669967,-.0331096){14}{\line(-1,0){.0669967}}
\multiput(80.634,121.252)(-.0611222,-.0336015){15}{\line(-1,0){.0611222}}
\multiput(79.717,120.748)(-.0525862,-.0319736){17}{\line(-1,0){.0525862}}
\multiput(78.823,120.205)(-.0482997,-.0323358){18}{\line(-1,0){.0482997}}
\multiput(77.954,119.623)(-.0443772,-.0326014){19}{\line(-1,0){.0443772}}
\multiput(77.11,119.003)(-.0407667,-.0327816){20}{\line(-1,0){.0407667}}
\multiput(76.295,118.348)(-.037426,-.032885){21}{\line(-1,0){.037426}}
\multiput(75.509,117.657)(-.034321,-.0329193){22}{\line(-1,0){.034321}}
\multiput(74.754,116.933)(-.0328518,-.0343856){22}{\line(0,-1){.0343856}}
\multiput(74.031,116.176)(-.0328115,-.0374906){21}{\line(0,-1){.0374906}}
\multiput(73.342,115.389)(-.0327014,-.040831){20}{\line(0,-1){.040831}}
\multiput(72.688,114.572)(-.0325142,-.0444412){19}{\line(0,-1){.0444412}}
\multiput(72.07,113.728)(-.0322408,-.0483631){18}{\line(0,-1){.0483631}}
\multiput(71.49,112.857)(-.0318702,-.0526489){17}{\line(0,-1){.0526489}}
\multiput(70.948,111.962)(-.0334814,-.0611881){15}{\line(0,-1){.0611881}}
\multiput(70.446,111.045)(-.0329779,-.0670616){14}{\line(0,-1){.0670616}}
\multiput(69.984,110.106)(-.0323293,-.0737012){13}{\line(0,-1){.0737012}}
\multiput(69.564,109.148)(-.031506,-.081295){12}{\line(0,-1){.081295}}
\multiput(69.186,108.172)(-.033514,-.099111){10}{\line(0,-1){.099111}}
\multiput(68.851,107.181)(-.032397,-.111644){9}{\line(0,-1){.111644}}
\multiput(68.559,106.176)(-.030931,-.12707){8}{\line(0,-1){.12707}}
\multiput(68.312,105.16)(-.028979,-.146627){7}{\line(0,-1){.146627}}
\multiput(68.109,104.133)(-.031574,-.206852){5}{\line(0,-1){.206852}}
\put(67.951,103.099){\line(0,-1){1.0402}}
\put(67.839,102.059){\line(0,-1){1.0441}}
\put(67.771,101.015){\line(0,-1){3.1359}}
\put(67.844,97.879){\line(0,-1){1.0399}}
\multiput(67.959,96.839)(.032116,-.206769){5}{\line(0,-1){.206769}}
\multiput(68.12,95.805)(.029363,-.14655){7}{\line(0,-1){.14655}}
\multiput(68.325,94.779)(.031264,-.126988){8}{\line(0,-1){.126988}}
\multiput(68.576,93.763)(.032689,-.111558){9}{\line(0,-1){.111558}}
\multiput(68.87,92.759)(.030703,-.090021){11}{\line(0,-1){.090021}}
\multiput(69.208,91.769)(.031719,-.081212){12}{\line(0,-1){.081212}}
\multiput(69.588,90.794)(.0325223,-.0736162){13}{\line(0,-1){.0736162}}
\multiput(70.011,89.837)(.0331534,-.066975){14}{\line(0,-1){.066975}}
\multiput(70.475,88.9)(.0336415,-.0611002){15}{\line(0,-1){.0611002}}
\multiput(70.98,87.983)(.032008,-.0525652){17}{\line(0,-1){.0525652}}
\multiput(71.524,87.09)(.0323674,-.0482785){18}{\line(0,-1){.0482785}}
\multiput(72.106,86.221)(.0326305,-.0443559){19}{\line(0,-1){.0443559}}
\multiput(72.726,85.378)(.0328082,-.0407452){20}{\line(0,-1){.0407452}}
\multiput(73.383,84.563)(.0329095,-.0374045){21}{\line(0,-1){.0374045}}
\multiput(74.074,83.778)(.0329418,-.0342995){22}{\line(0,-1){.0342995}}
\multiput(74.798,83.023)(.0344071,-.0328293){22}{\line(1,0){.0344071}}
\multiput(75.555,82.301)(.037512,-.0327869){21}{\line(1,0){.037512}}
\multiput(76.343,81.612)(.0408524,-.0326747){20}{\line(1,0){.0408524}}
\multiput(77.16,80.959)(.0444625,-.0324851){19}{\line(1,0){.0444625}}
\multiput(78.005,80.341)(.0483842,-.0322091){18}{\line(1,0){.0483842}}
\multiput(78.876,79.762)(.0526697,-.0318357){17}{\line(1,0){.0526697}}
\multiput(79.771,79.22)(.06121,-.0334413){15}{\line(1,0){.06121}}
\multiput(80.689,78.719)(.0670832,-.032934){14}{\line(1,0){.0670832}}
\multiput(81.629,78.258)(.0737223,-.0322811){13}{\line(1,0){.0737223}}
\multiput(82.587,77.838)(.081316,-.031453){12}{\line(1,0){.081316}}
\multiput(83.563,77.461)(.099133,-.033449){10}{\line(1,0){.099133}}
\multiput(84.554,77.126)(.111665,-.032324){9}{\line(1,0){.111665}}
\multiput(85.559,76.835)(.12709,-.030848){8}{\line(1,0){.12709}}
\multiput(86.576,76.589)(.171087,-.033697){6}{\line(1,0){.171087}}
\multiput(87.602,76.386)(.206873,-.031438){5}{\line(1,0){.206873}}
\put(88.637,76.229){\line(1,0){1.0402}}
\put(89.677,76.117){\line(1,0){1.0441}}
\put(90.721,76.051){\line(1,0){2.092}}
\put(92.813,76.055){\line(1,0){1.0439}}
\put(93.857,76.125){\line(1,0){1.0398}}
\multiput(94.897,76.241)(.206748,.032251){5}{\line(1,0){.206748}}
\multiput(95.93,76.403)(.146531,.029459){7}{\line(1,0){.146531}}
\multiput(96.956,76.609)(.126968,.031347){8}{\line(1,0){.126968}}
\multiput(97.972,76.86)(.111537,.032762){9}{\line(1,0){.111537}}
\multiput(98.976,77.155)(.090001,.030762){11}{\line(1,0){.090001}}
\multiput(99.966,77.493)(.081192,.031772){12}{\line(1,0){.081192}}
\multiput(100.94,77.874)(.0735949,.0325705){13}{\line(1,0){.0735949}}
\multiput(101.897,78.298)(.0669533,.0331973){14}{\line(1,0){.0669533}}
\multiput(102.834,78.762)(.0610781,.0336815){15}{\line(1,0){.0610781}}
\multiput(103.75,79.268)(.0525443,.0320424){17}{\line(1,0){.0525443}}
\multiput(104.644,79.812)(.0482573,.032399){18}{\line(1,0){.0482573}}
\multiput(105.512,80.395)(.0443345,.0326595){19}{\line(1,0){.0443345}}
\multiput(106.355,81.016)(.0407237,.0328349){20}{\line(1,0){.0407237}}
\multiput(107.169,81.673)(.0373829,.032934){21}{\line(1,0){.0373829}}
\multiput(107.954,82.364)(.0342779,.0329642){22}{\line(1,0){.0342779}}
\multiput(108.708,83.09)(.0328068,.0344286){22}{\line(0,1){.0344286}}
\multiput(109.43,83.847)(.0327623,.0375335){21}{\line(0,1){.0375335}}
\multiput(110.118,84.635)(.0326479,.0408738){20}{\line(0,1){.0408738}}
\multiput(110.771,85.453)(.032456,.0444837){19}{\line(0,1){.0444837}}
\multiput(111.388,86.298)(.0321775,.0484053){18}{\line(0,1){.0484053}}
\multiput(111.967,87.169)(.0318013,.0526906){17}{\line(0,1){.0526906}}
\multiput(112.507,88.065)(.0334012,.0612319){15}{\line(0,1){.0612319}}
\multiput(113.008,88.983)(.0328901,.0671047){14}{\line(0,1){.0671047}}
\multiput(113.469,89.923)(.0322328,.0737434){13}{\line(0,1){.0737434}}
\multiput(113.888,90.881)(.031399,.081336){12}{\line(0,1){.081336}}
\multiput(114.265,91.858)(.033384,.099155){10}{\line(0,1){.099155}}
\multiput(114.598,92.849)(.032251,.111686){9}{\line(0,1){.111686}}
\multiput(114.889,93.854)(.030765,.12711){8}{\line(0,1){.12711}}
\multiput(115.135,94.871)(.033585,.171109){6}{\line(0,1){.171109}}
\multiput(115.336,95.898)(.031303,.206894){5}{\line(0,1){.206894}}
\put(115.493,96.932){\line(0,1){1.0403}}
\put(115.604,97.973){\line(0,1){2.0274}}
\put(115.263,100){\line(0,1){.8666}}
\put(115.242,100.867){\line(0,1){.8646}}
\put(115.181,101.731){\line(0,1){.8607}}
\multiput(115.078,102.592)(-.028623,.170984){5}{\line(0,1){.170984}}
\multiput(114.935,103.447)(-.030575,.141197){6}{\line(0,1){.141197}}
\multiput(114.752,104.294)(-.031911,.119648){7}{\line(0,1){.119648}}
\multiput(114.528,105.132)(-.032851,.103252){8}{\line(0,1){.103252}}
\multiput(114.266,105.958)(-.033515,.090293){9}{\line(0,1){.090293}}
\multiput(113.964,106.77)(-.030891,.072494){11}{\line(0,1){.072494}}
\multiput(113.624,107.568)(-.031433,.065037){12}{\line(0,1){.065037}}
\multiput(113.247,108.348)(-.0318261,.0585924){13}{\line(0,1){.0585924}}
\multiput(112.833,109.11)(-.032097,.0529462){14}{\line(0,1){.0529462}}
\multiput(112.384,109.851)(-.0322645,.0479418){15}{\line(0,1){.0479418}}
\multiput(111.9,110.57)(-.0323431,.0434621){16}{\line(0,1){.0434621}}
\multiput(111.383,111.266)(-.0323442,.0394176){17}{\line(0,1){.0394176}}
\multiput(110.833,111.936)(-.0322765,.0357388){18}{\line(0,1){.0357388}}
\multiput(110.252,112.579)(-.0321473,.0323713){19}{\line(0,1){.0323713}}
\multiput(109.641,113.194)(-.0355139,.0325239){18}{\line(-1,0){.0355139}}
\multiput(109.002,113.779)(-.0391921,.0326171){17}{\line(-1,0){.0391921}}
\multiput(108.335,114.334)(-.0432365,.0326441){16}{\line(-1,0){.0432365}}
\multiput(107.644,114.856)(-.0477167,.0325965){15}{\line(-1,0){.0477167}}
\multiput(106.928,115.345)(-.0527221,.0324638){14}{\line(-1,0){.0527221}}
\multiput(106.19,115.8)(-.05837,.0322321){13}{\line(-1,0){.05837}}
\multiput(105.431,116.219)(-.064817,.031883){12}{\line(-1,0){.064817}}
\multiput(104.653,116.601)(-.072278,.031393){11}{\line(-1,0){.072278}}
\multiput(103.858,116.947)(-.081053,.030727){10}{\line(-1,0){.081053}}
\multiput(103.048,117.254)(-.103021,.033567){8}{\line(-1,0){.103021}}
\multiput(102.223,117.522)(-.119424,.032741){7}{\line(-1,0){.119424}}
\multiput(101.387,117.752)(-.140981,.031555){6}{\line(-1,0){.140981}}
\multiput(100.541,117.941)(-.170781,.029809){5}{\line(-1,0){.170781}}
\put(99.688,118.09){\line(-1,0){.86}}
\put(98.828,118.198){\line(-1,0){.8642}}
\put(97.963,118.266){\line(-1,0){.8664}}
\put(97.097,118.293){\line(-1,0){.8667}}
\put(96.23,118.278){\line(-1,0){.865}}
\put(95.365,118.222){\line(-1,0){.8614}}
\multiput(94.504,118.126)(-.171178,-.027435){5}{\line(-1,0){.171178}}
\multiput(93.648,117.989)(-.141405,-.029594){6}{\line(-1,0){.141405}}
\multiput(92.8,117.811)(-.119867,-.03108){7}{\line(-1,0){.119867}}
\multiput(91.96,117.594)(-.103478,-.032133){8}{\line(-1,0){.103478}}
\multiput(91.133,117.337)(-.090524,-.032888){9}{\line(-1,0){.090524}}
\multiput(90.318,117.041)(-.079978,-.033425){10}{\line(-1,0){.079978}}
\multiput(89.518,116.706)(-.065254,-.03098){12}{\line(-1,0){.065254}}
\multiput(88.735,116.335)(-.0588119,-.0314185){13}{\line(-1,0){.0588119}}
\multiput(87.971,115.926)(-.0531677,-.0317286){14}{\line(-1,0){.0531677}}
\multiput(87.226,115.482)(-.0481647,-.0319309){15}{\line(-1,0){.0481647}}
\multiput(86.504,115.003)(-.0436856,-.0320406){16}{\line(-1,0){.0436856}}
\multiput(85.805,114.49)(-.0396412,-.0320697){17}{\line(-1,0){.0396412}}
\multiput(85.131,113.945)(-.0359621,-.0320276){18}{\line(-1,0){.0359621}}
\multiput(84.484,113.369)(-.0344045,-.0336953){18}{\line(-1,0){.0344045}}
\multiput(83.864,112.762)(-.0327696,-.0352872){18}{\line(0,-1){.0352872}}
\multiput(83.274,112.127)(-.0328884,-.0389647){17}{\line(0,-1){.0389647}}
\multiput(82.715,111.465)(-.0329435,-.0430089){16}{\line(0,-1){.0430089}}
\multiput(82.188,110.776)(-.032927,-.0474892){15}{\line(0,-1){.0474892}}
\multiput(81.694,110.064)(-.032829,-.0524954){14}{\line(0,-1){.0524954}}
\multiput(81.235,109.329)(-.0326365,-.0581449){13}{\line(0,-1){.0581449}}
\multiput(80.81,108.573)(-.032333,-.064594){12}{\line(0,-1){.064594}}
\multiput(80.422,107.798)(-.031894,-.072058){11}{\line(0,-1){.072058}}
\multiput(80.072,107.006)(-.031289,-.080837){10}{\line(0,-1){.080837}}
\multiput(79.759,106.197)(-.030472,-.091365){9}{\line(0,-1){.091365}}
\multiput(79.484,105.375)(-.03357,-.119194){7}{\line(0,-1){.119194}}
\multiput(79.249,104.541)(-.032533,-.140758){6}{\line(0,-1){.140758}}
\multiput(79.054,103.696)(-.030994,-.17057){5}{\line(0,-1){.17057}}
\put(78.899,102.843){\line(0,-1){.8592}}
\put(78.785,101.984){\line(0,-1){.8637}}
\put(78.711,101.12){\line(0,-1){1.733}}
\put(78.687,99.387){\line(0,-1){.8654}}
\put(78.737,98.522){\line(0,-1){.8621}}
\multiput(78.827,97.66)(.03281,-.21421){4}{\line(0,-1){.21421}}
\multiput(78.959,96.803)(.028612,-.141608){6}{\line(0,-1){.141608}}
\multiput(79.13,95.953)(.030247,-.12008){7}{\line(0,-1){.12008}}
\multiput(79.342,95.113)(.031414,-.103698){8}{\line(0,-1){.103698}}
\multiput(79.593,94.283)(.032259,-.09075){9}{\line(0,-1){.09075}}
\multiput(79.884,93.466)(.032869,-.080208){10}{\line(0,-1){.080208}}
\multiput(80.212,92.664)(.033302,-.071419){11}{\line(0,-1){.071419}}
\multiput(80.579,91.879)(.033594,-.063948){12}{\line(0,-1){.063948}}
\multiput(80.982,91.111)(.0313587,-.0533867){14}{\line(0,-1){.0533867}}
\multiput(81.421,90.364)(.0315957,-.0483852){15}{\line(0,-1){.0483852}}
\multiput(81.895,89.638)(.0317366,-.043907){16}{\line(0,-1){.043907}}
\multiput(82.402,88.936)(.0317938,-.0398629){17}{\line(0,-1){.0398629}}
\multiput(82.943,88.258)(.0336464,-.038312){17}{\line(0,-1){.038312}}
\multiput(83.515,87.607)(.0334556,-.0346376){18}{\line(0,-1){.0346376}}
\multiput(84.117,86.983)(.0350589,-.0330138){18}{\line(1,0){.0350589}}
\multiput(84.748,86.389)(.0387354,-.0331581){17}{\line(1,0){.0387354}}
\multiput(85.407,85.825)(.0427791,-.0332413){16}{\line(1,0){.0427791}}
\multiput(86.091,85.293)(.0472595,-.0332559){15}{\line(1,0){.0472595}}
\multiput(86.8,84.795)(.0522662,-.0331927){14}{\line(1,0){.0522662}}
\multiput(87.532,84.33)(.0579169,-.0330394){13}{\line(1,0){.0579169}}
\multiput(88.285,83.9)(.064368,-.03278){12}{\line(1,0){.064368}}
\multiput(89.057,83.507)(.071835,-.032394){11}{\line(1,0){.071835}}
\multiput(89.847,83.151)(.080618,-.03185){10}{\line(1,0){.080618}}
\multiput(90.653,82.832)(.091152,-.031106){9}{\line(1,0){.091152}}
\multiput(91.474,82.552)(.104088,-.030097){8}{\line(1,0){.104088}}
\multiput(92.307,82.311)(.140529,-.033509){6}{\line(1,0){.140529}}
\multiput(93.15,82.11)(.170351,-.032178){5}{\line(1,0){.170351}}
\multiput(94.001,81.949)(.2146,-.03009){4}{\line(1,0){.2146}}
\put(94.86,81.829){\line(1,0){.8632}}
\put(95.723,81.75){\line(1,0){.866}}
\put(96.589,81.711){\line(1,0){.8668}}
\put(97.456,81.713){\line(1,0){.8657}}
\put(98.322,81.757){\line(1,0){.8627}}
\multiput(99.184,81.842)(.21443,.03132){4}{\line(1,0){.21443}}
\multiput(100.042,81.967)(.170163,.033154){5}{\line(1,0){.170163}}
\multiput(100.893,82.133)(.120287,.029413){7}{\line(1,0){.120287}}
\multiput(101.735,82.338)(.103914,.030693){8}{\line(1,0){.103914}}
\multiput(102.566,82.584)(.090972,.031628){9}{\line(1,0){.090972}}
\multiput(103.385,82.869)(.080434,.032312){10}{\line(1,0){.080434}}
\multiput(104.189,83.192)(.071648,.032805){11}{\line(1,0){.071648}}
\multiput(104.977,83.553)(.064179,.033149){12}{\line(1,0){.064179}}
\multiput(105.747,83.95)(.0577265,.033371){13}{\line(1,0){.0577265}}
\multiput(106.498,84.384)(.0520751,.0334918){14}{\line(1,0){.0520751}}
\multiput(107.227,84.853)(.047068,.0335264){15}{\line(1,0){.047068}}
\multiput(107.933,85.356)(.0425878,.033486){16}{\line(1,0){.0425878}}
\multiput(108.614,85.892)(.0385446,.0333797){17}{\line(1,0){.0385446}}
\multiput(109.27,86.459)(.034869,.0332143){18}{\line(1,0){.034869}}
\multiput(109.897,87.057)(.0332564,.0348288){18}{\line(0,1){.0348288}}
\multiput(110.496,87.684)(.0334262,.0385043){17}{\line(0,1){.0385043}}
\multiput(111.064,88.339)(.0335375,.0425473){16}{\line(0,1){.0425473}}
\multiput(111.601,89.019)(.0335832,.0470275){15}{\line(0,1){.0470275}}
\multiput(112.104,89.725)(.0335547,.0520345){14}{\line(0,1){.0520345}}
\multiput(112.574,90.453)(.0334407,.0576861){13}{\line(0,1){.0576861}}
\multiput(113.009,91.203)(.033226,.064139){12}{\line(0,1){.064139}}
\multiput(113.408,91.973)(.032892,.071609){11}{\line(0,1){.071609}}
\multiput(113.77,92.761)(.032409,.080395){10}{\line(0,1){.080395}}
\multiput(114.094,93.564)(.031738,.090933){9}{\line(0,1){.090933}}
\multiput(114.379,94.383)(.030819,.103877){8}{\line(0,1){.103877}}
\multiput(114.626,95.214)(.029558,.120251){7}{\line(0,1){.120251}}
\multiput(114.833,96.056)(.03336,.170123){5}{\line(0,1){.170123}}
\multiput(114.999,96.906)(.03158,.21439){4}{\line(0,1){.21439}}
\put(115.126,97.764){\line(0,1){.8626}}
\put(115.211,98.626){\line(0,1){1.3736}}
\put(115.47,100.5){\circle*{1.118}}
\put(73.72,100.25){\circle*{.5}}
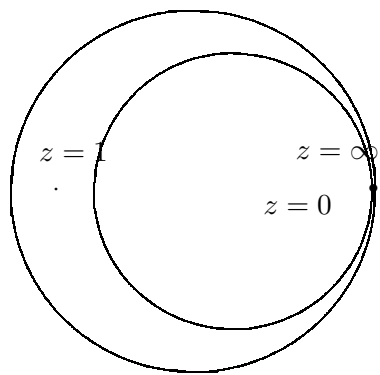
\put(70.22,68){\makebox(0,0)[cc]{Fig.1 Spectral curve for CS system}}
\put(71.5,104){$z=1$}
\put(105.25,104.25){$z=\infty$}
\put(101,97){$z=0$}
\end{picture}

 \vspace{-55mm}
\begin{picture}(108.75,120.25)(0,0)
\put(86.75,97.75){\oval(16,36.5)[]}
\put(104.766,97.5){\line(0,1){.857}}
\put(104.745,98.357){\line(0,1){.8551}}
\put(104.684,99.212){\line(0,1){.8512}}
\multiput(104.582,100.063)(-.028425,.169079){5}{\line(0,1){.169079}}
\multiput(104.44,100.909)(-.030363,.139613){6}{\line(0,1){.139613}}
\multiput(104.258,101.746)(-.031689,.118294){7}{\line(0,1){.118294}}
\multiput(104.036,102.574)(-.03262,.102071){8}{\line(0,1){.102071}}
\multiput(103.775,103.391)(-.033279,.089248){9}{\line(0,1){.089248}}
\multiput(103.476,104.194)(-.033738,.078808){10}{\line(0,1){.078808}}
\multiput(103.138,104.982)(-.031208,.064261){12}{\line(0,1){.064261}}
\multiput(102.764,105.753)(-.0315962,.0578804){13}{\line(0,1){.0578804}}
\multiput(102.353,106.506)(-.0318629,.0522896){14}{\line(0,1){.0522896}}
\multiput(101.907,107.238)(-.0320266,.0473337){15}{\line(0,1){.0473337}}
\multiput(101.427,107.948)(-.0321019,.0428968){16}{\line(0,1){.0428968}}
\multiput(100.913,108.634)(-.0321,.0388904){17}{\line(0,1){.0388904}}
\multiput(100.367,109.295)(-.0320296,.0352461){18}{\line(0,1){.0352461}}
\multiput(99.791,109.93)(-.03367,.0336825){18}{\line(0,1){.0336825}}
\multiput(99.185,110.536)(-.0352342,.0320427){18}{\line(-1,0){.0352342}}
\multiput(98.55,111.113)(-.0388785,.0321144){17}{\line(-1,0){.0388785}}
\multiput(97.89,111.659)(-.0428848,.0321179){16}{\line(-1,0){.0428848}}
\multiput(97.203,112.173)(-.0473218,.0320442){15}{\line(-1,0){.0473218}}
\multiput(96.494,112.653)(-.0522777,.0318823){14}{\line(-1,0){.0522777}}
\multiput(95.762,113.1)(-.0578687,.0316177){13}{\line(-1,0){.0578687}}
\multiput(95.009,113.511)(-.06425,.031232){12}{\line(-1,0){.06425}}
\multiput(94.238,113.886)(-.071632,.030698){11}{\line(-1,0){.071632}}
\multiput(93.45,114.223)(-.089236,.033312){9}{\line(-1,0){.089236}}
\multiput(92.647,114.523)(-.102059,.032658){8}{\line(-1,0){.102059}}
\multiput(91.831,114.784)(-.118283,.031733){7}{\line(-1,0){.118283}}
\multiput(91.003,115.006)(-.139601,.030415){6}{\line(-1,0){.139601}}
\multiput(90.165,115.189)(-.169068,.028488){5}{\line(-1,0){.169068}}
\put(89.32,115.331){\line(-1,0){.8512}}
\put(88.469,115.433){\line(-1,0){.8551}}
\put(87.614,115.495){\line(-1,0){1.714}}
\put(85.9,115.496){\line(-1,0){.8551}}
\put(85.045,115.435){\line(-1,0){.8512}}
\multiput(84.193,115.333)(-.169089,-.028362){5}{\line(-1,0){.169089}}
\multiput(83.348,115.191)(-.139624,-.030312){6}{\line(-1,0){.139624}}
\multiput(82.51,115.01)(-.118306,-.031645){7}{\line(-1,0){.118306}}
\multiput(81.682,114.788)(-.102084,-.032583){8}{\line(-1,0){.102084}}
\multiput(80.865,114.527)(-.08926,-.033246){9}{\line(-1,0){.08926}}
\multiput(80.062,114.228)(-.07882,-.033709){10}{\line(-1,0){.07882}}
\multiput(79.274,113.891)(-.064273,-.031184){12}{\line(-1,0){.064273}}
\multiput(78.503,113.517)(-.0578922,-.0315747){13}{\line(-1,0){.0578922}}
\multiput(77.75,113.106)(-.0523014,-.0318434){14}{\line(-1,0){.0523014}}
\multiput(77.018,112.661)(-.0473456,-.032009){15}{\line(-1,0){.0473456}}
\multiput(76.308,112.181)(-.0429087,-.032086){16}{\line(-1,0){.0429087}}
\multiput(75.621,111.667)(-.0389024,-.0320855){17}{\line(-1,0){.0389024}}
\multiput(74.96,111.122)(-.035258,-.0320165){18}{\line(-1,0){.035258}}
\multiput(74.325,110.545)(-.033695,-.0336575){18}{\line(-1,0){.033695}}
\multiput(73.718,109.94)(-.0320558,-.0352223){18}{\line(0,-1){.0352223}}
\multiput(73.141,109.306)(-.0321289,-.0388666){17}{\line(0,-1){.0388666}}
\multiput(72.595,108.645)(-.0321338,-.0428729){16}{\line(0,-1){.0428729}}
\multiput(72.081,107.959)(-.0320618,-.0473098){15}{\line(0,-1){.0473098}}
\multiput(71.6,107.249)(-.0319017,-.0522659){14}{\line(0,-1){.0522659}}
\multiput(71.154,106.517)(-.0316392,-.0578569){13}{\line(0,-1){.0578569}}
\multiput(70.742,105.765)(-.031255,-.064238){12}{\line(0,-1){.064238}}
\multiput(70.367,104.994)(-.030725,-.07162){11}{\line(0,-1){.07162}}
\multiput(70.029,104.207)(-.033346,-.089223){9}{\line(0,-1){.089223}}
\multiput(69.729,103.404)(-.032696,-.102047){8}{\line(0,-1){.102047}}
\multiput(69.468,102.587)(-.031777,-.118271){7}{\line(0,-1){.118271}}
\multiput(69.245,101.759)(-.030467,-.13959){6}{\line(0,-1){.13959}}
\multiput(69.062,100.922)(-.028551,-.169058){5}{\line(0,-1){.169058}}
\put(68.92,100.077){\line(0,-1){.8511}}
\put(68.817,99.225){\line(0,-1){.855}}
\put(68.755,98.37){\line(0,-1){2.5691}}
\put(68.815,95.801){\line(0,-1){.8513}}
\multiput(68.916,94.95)(.028299,-.1691){5}{\line(0,-1){.1691}}
\multiput(69.057,94.104)(.03026,-.139635){6}{\line(0,-1){.139635}}
\multiput(69.239,93.267)(.031601,-.118318){7}{\line(0,-1){.118318}}
\multiput(69.46,92.438)(.032545,-.102096){8}{\line(0,-1){.102096}}
\multiput(69.72,91.622)(.033213,-.089273){9}{\line(0,-1){.089273}}
\multiput(70.019,90.818)(.03368,-.078833){10}{\line(0,-1){.078833}}
\multiput(70.356,90.03)(.03116,-.064284){12}{\line(0,-1){.064284}}
\multiput(70.73,89.258)(.0315532,-.0579039){13}{\line(0,-1){.0579039}}
\multiput(71.14,88.506)(.031824,-.0523132){14}{\line(0,-1){.0523132}}
\multiput(71.586,87.773)(.0319914,-.0473575){15}{\line(0,-1){.0473575}}
\multiput(72.066,87.063)(.03207,-.0429206){16}{\line(0,-1){.0429206}}
\multiput(72.579,86.376)(.0320711,-.0389143){17}{\line(0,-1){.0389143}}
\multiput(73.124,85.715)(.0320034,-.0352699){18}{\line(0,-1){.0352699}}
\multiput(73.7,85.08)(.033645,-.0337075){18}{\line(0,-1){.0337075}}
\multiput(74.306,84.473)(.0352104,-.0320689){18}{\line(1,0){.0352104}}
\multiput(74.939,83.896)(.0388546,-.0321433){17}{\line(1,0){.0388546}}
\multiput(75.6,83.349)(.0428609,-.0321497){16}{\line(1,0){.0428609}}
\multiput(76.286,82.835)(.0472979,-.0320794){15}{\line(1,0){.0472979}}
\multiput(76.995,82.354)(.052254,-.0319211){14}{\line(1,0){.052254}}
\multiput(77.727,81.907)(.0578452,-.0316607){13}{\line(1,0){.0578452}}
\multiput(78.479,81.495)(.064226,-.031279){12}{\line(1,0){.064226}}
\multiput(79.249,81.12)(.071609,-.030751){11}{\line(1,0){.071609}}
\multiput(80.037,80.782)(.089211,-.033379){9}{\line(1,0){.089211}}
\multiput(80.84,80.481)(.102035,-.032734){8}{\line(1,0){.102035}}
\multiput(81.656,80.219)(.118259,-.031821){7}{\line(1,0){.118259}}
\multiput(82.484,79.997)(.139579,-.030519){6}{\line(1,0){.139579}}
\multiput(83.322,79.814)(.169047,-.028614){5}{\line(1,0){.169047}}
\put(84.167,79.671){\line(1,0){.8511}}
\put(85.018,79.568){\line(1,0){.855}}
\put(85.873,79.506){\line(1,0){1.714}}
\put(87.587,79.504){\line(1,0){.8551}}
\put(88.442,79.564){\line(1,0){.8513}}
\multiput(89.293,79.665)(.16911,.028237){5}{\line(1,0){.16911}}
\multiput(90.139,79.806)(.139646,.030208){6}{\line(1,0){.139646}}
\multiput(90.977,79.987)(.11833,.031557){7}{\line(1,0){.11833}}
\multiput(91.805,80.208)(.102108,.032507){8}{\line(1,0){.102108}}
\multiput(92.622,80.468)(.089285,.03318){9}{\line(1,0){.089285}}
\multiput(93.426,80.767)(.078845,.033651){10}{\line(1,0){.078845}}
\multiput(94.214,81.103)(.064296,.031136){12}{\line(1,0){.064296}}
\multiput(94.986,81.477)(.0579156,.0315317){13}{\line(1,0){.0579156}}
\multiput(95.738,81.887)(.0523251,.0318046){14}{\line(1,0){.0523251}}
\multiput(96.471,82.332)(.0473693,.0319738){15}{\line(1,0){.0473693}}
\multiput(97.182,82.812)(.0429325,.0320541){16}{\line(1,0){.0429325}}
\multiput(97.868,83.325)(.0389262,.0320566){17}{\line(1,0){.0389262}}
\multiput(98.53,83.87)(.0352818,.0319903){18}{\line(1,0){.0352818}}
\multiput(99.165,84.445)(.03372,.0336324){18}{\line(1,0){.03372}}
\multiput(99.772,85.051)(.0320819,.0351984){18}{\line(0,1){.0351984}}
\multiput(100.35,85.684)(.0321577,.0388427){17}{\line(0,1){.0388427}}
\multiput(100.896,86.345)(.0321657,.042849){16}{\line(0,1){.042849}}
\multiput(101.411,87.03)(.032097,.047286){15}{\line(0,1){.047286}}
\multiput(101.893,87.74)(.0319406,.0522422){14}{\line(0,1){.0522422}}
\multiput(102.34,88.471)(.0316822,.0578334){13}{\line(0,1){.0578334}}
\multiput(102.752,89.223)(.031303,.064215){12}{\line(0,1){.064215}}
\multiput(103.127,89.993)(.030778,.071598){11}{\line(0,1){.071598}}
\multiput(103.466,90.781)(.033412,.089199){9}{\line(0,1){.089199}}
\multiput(103.766,91.584)(.032772,.102023){8}{\line(0,1){.102023}}
\multiput(104.029,92.4)(.031865,.118247){7}{\line(0,1){.118247}}
\multiput(104.252,93.228)(.030571,.139567){6}{\line(0,1){.139567}}
\multiput(104.435,94.065)(.028676,.169036){5}{\line(0,1){.169036}}
\put(104.579,94.91){\line(0,1){.851}}
\put(104.682,95.761){\line(0,1){.855}}
\put(104.744,96.616){\line(0,1){.8838}}
\put(86.75,116){\circle*{1.118}}
\put(87,79.75){\circle*{1}}
\put(99.5,97.75){\circle*{1.118}}
\put(74.25,97.25){\circle*{.707}}
\put(81,120.25){$z_1=z_2=\infty$}
\put(81.5,75.5){$z_1=z_2=0$}
\put(61.5,98){$z_1=1$}
\put(108.75,98.75){$z_2=1$}
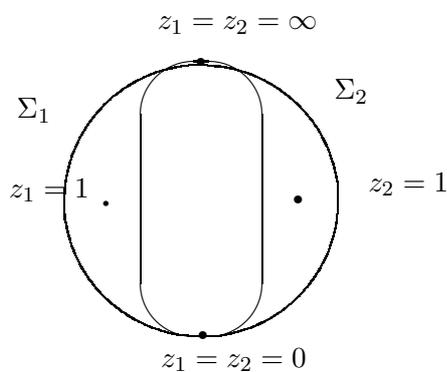
\put(69.5,70.5){\makebox(0,0)[cc]{Fig.2 Spectral curve for generalized CS system}}
\put(62.5,108.5){$\Sigma_1$}
\put(104.25,111){$\Sigma_2$}
\end{picture}


\vspace{-70mm}

\begin{small}

\end{small}

\end{document}